\documentclass[mathleft,fleqn,%
]{an}
\usepackage{graphicx}
\usepackage[varg]{txfonts}
\usepackage{natbib}
\bibpunct{(}{)}{;}{a}{}{,}
\begin{document}

\Pagespan{1}{}
\Yearpublication{2016}%
\Yearsubmission{2016}%
\Month{0}%
\Volume{999}%
\Issue{0}%
\DOI{asna.201400000}%

\title{Kinematical properties of coronal mass ejections}

\author{M.\,Temmer\inst{1}\fnmsep\thanks{Corresponding author:
        {manuela.temmer@uni-graz.at}}
}
\titlerunning{Kinematical properties of coronal mass ejections}
\authorrunning{M.\,Temmer}
\institute{Institute of Physics, University of Graz, Austria
}

\received{XXXX}
\accepted{XXXX}
\publonline{XXXX}


\abstract{%
Coronal mass ejections (CMEs) are the most dynamic phenomena in our solar system. They abruptly disrupt the continuous outflow of solar wind by expelling huge clouds of magnetized plasma into interplanetary space with velocities enabling to cross the Sun-Earth distance within a few days. Earth-directed CMEs may cause severe geomagnetic storms when their embedded magnetic fields and the shocks ahead compress and reconnect with the Earth's magnetic field. The transit times and impacts in detail depend on the initial CME velocity, size, and mass, as well as on the conditions and coupling processes with the ambient solar wind flow in interplanetary space. The observed CME parameters may be severly affected by projection effects and the constant changing environmental conditions are hard to derive. This makes it difficult to fully understand the physics behind CME evolution, preventing to do a reliable forecast of Earth-directed events. This short review focusing on observational data, shows recent methods which were developed to derive the CME kinematical profile for the entire Sun-Earth distance range as well as studies which were performed to shed light on the physical processes that CMEs encounter when propagating from Sun to Earth.  }

\maketitle

\section{Introduction}
The initiation of coronal mass ejections (CMEs) can be described as disrupted equilibrium, which becomes somehow unstable and starts to erupt \citep[for a recent review on CME initiation models see e.g.][]{webb12}. Due to the successive stretching of magnetic field lines magnetic reconnection may set in, enabling fast as well as slow reconfiguration processes \citep{forbes00}. Slow magnetic reconfiguration processes presumably dominate high up in the corona launching so-called stealth CMEs \citep{robbrecht09}. The observed CME front is formed due to plasma-pileup and compression of plasma \citep[e.g.,][]{ontiveros09,vourlidas13}. With peak acceleration values of more than 600~m~s$^{-2}$ \citep[for statistics see e.g.,][]{bein11}, CMEs abruptly disrupt the continuous outflow of solar wind. Together with their embedded magnetic field (magnetized plasma clouds) and shocks ahead, CMEs propagate into interplanetary (IP) space with velocities of a few hundred to a few thousand km/s \citep[e.g.,][]{yashiro04} and may cause severe space weather effects when interacting with the Earth's magnetic field \citep[see e.g.,][and references therein]{schwenn06,pulkkinen07}. Even thermospheric responses in terms of a neutral density increase are observed due to CME-Earth interactions \cite[e.g.,][]{guo10, krauss15}. The transit times to reach a distance of 1~AU is of about one to five days, and depends on the CME's initial velocity, size, and mass, as well as on the conditions of the ambient solar wind flow in IP space \citep[see e.g.,][]{bothmer07}.

\begin{figure}
\includegraphics[width=\linewidth]{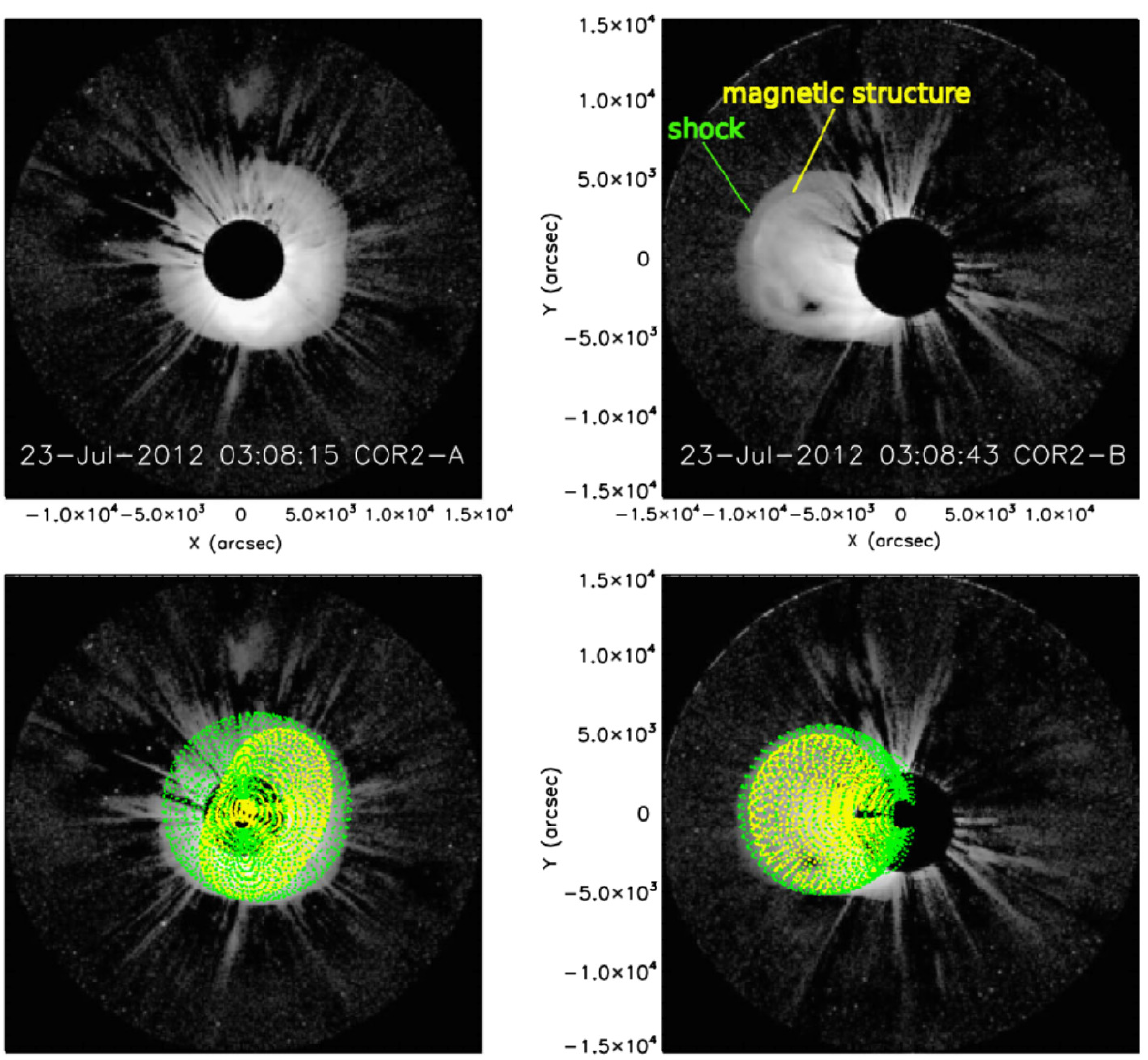}
\caption{Reconstructed CME (magnetic structure in yellow and shock in green) using the graduated cylindrical shell model \citep{thernisien06} applied on simultaneous observations from STEREO-A and STEREO-B spacecraft. Adapted from \cite{temmer15}.}
\label{f1}
\end{figure}

The measurement of characteristic CME parameters such as speed, width, propagation direction, are usually executed from single vantage points, hence, from observations projected onto the plane-of-sky \citep[e.g.,][]{hundhausen93}. Therefore, all derived values are severely affected by projection effects \citep[see e.g.,][]{burkepile04,cremades04}. Of special interest are halo CMEs, i.e.\,CMEs propagating towards the observer especially at Earth, that were subject to a large number of studies using the coronagraphs aboard the Solar and Heliospheric Observatory launched in 1995 \citep[SoHO;][]{brueckner95}. The question arose whether halo CMEs would be different compared to limb CMEs \citep{chen11}. Such questions can only be answered when using multipoint observational data, as from the Solar TErrestrial RElations Observatory launched in 2006 \citep[STEREO;][]{howard08}. In a recent study \cite{kwon15} showed that halo CMEs do not reflect the actual size of a CME but that of the fast shock wave around the magnetic structure (cf.\,Figure~\ref{f1}). Using 3D speeds from STEREO, the uncertainties in estimated arrival times of CMEs could be decreased by up to a factor of two compared to projected speeds as derived from single spacecraft data \citep[e.g.,][]{mishra13,colaninno13,shi15}. In this respect, STEREO observations opened a new observational window and enabled for the first time to study in detail the consequences of projection effects, and as such this could largely improve the understanding of CMEs. 

\section{Kinematic evolution of CMEs}
\begin{figure}
\includegraphics[width=\linewidth]{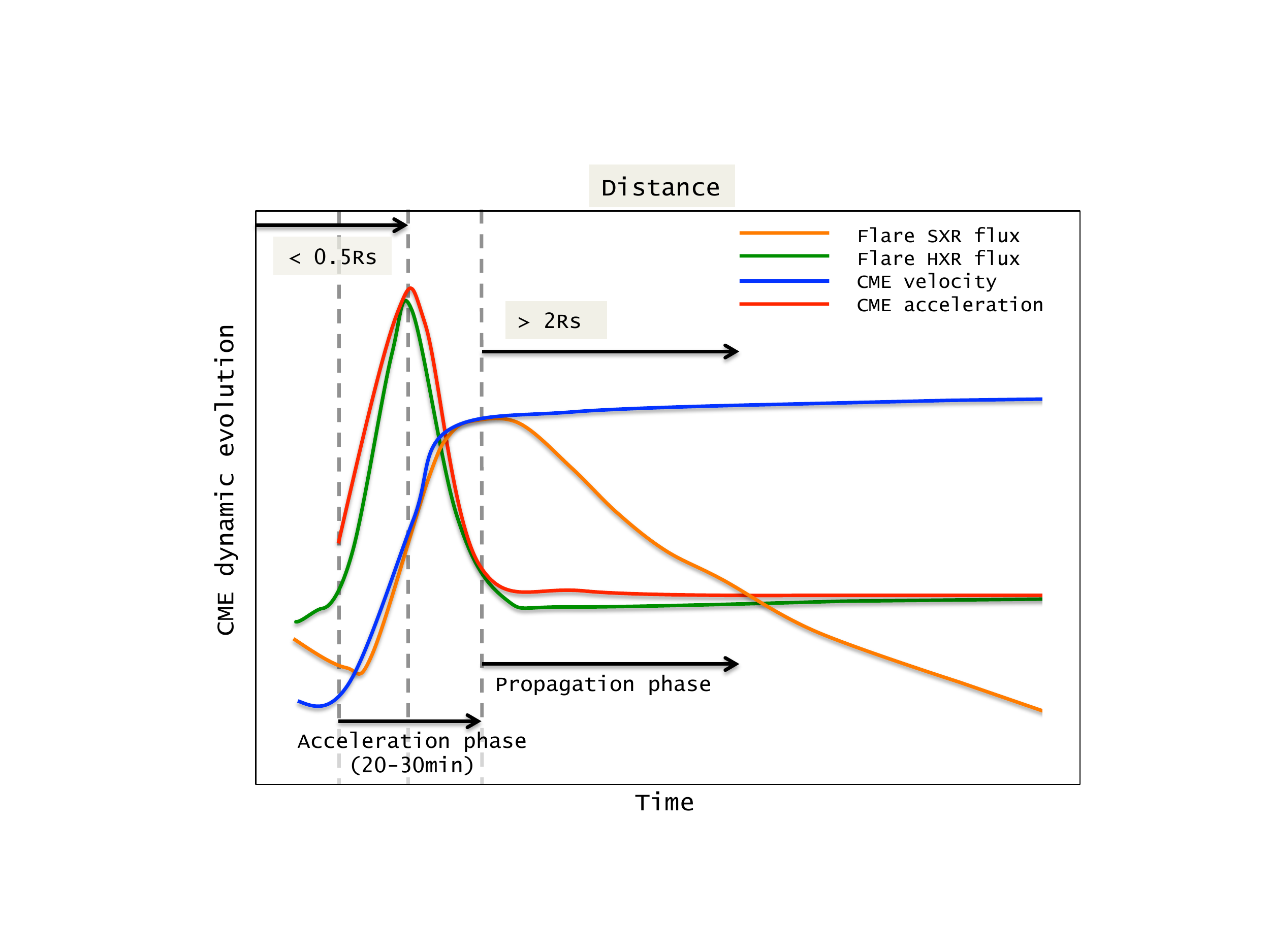}
\caption{Schematic view on the dynamic evolution of a CME with focus on the early evolution close to the Sun. The energy release of the associated flare as measured by the hard X-ray (HXR) flux is synchronized with the CME acceleration phase, the measured soft X-ray flux (SXR) goes along with the CME speed until its maximum.}
\label{f2}
\end{figure}

Close to the Sun, the propelling Lorentz force, as a consequence of magnetic reconnection \citep{chen89,chen96,kliem06}, triggers the CME. In IP space, it is the drag acceleration owing to the ambient solar wind flow that dominates the propagation behaviour of a CME \citep[e.g.,][]{vrsnak90,cargill96,chen96}. Figure~\ref{f2} shows a schematic view on the evolution of a CME \citep[see also Figure~1 in][]{zhang06}. According to  studies by \cite{zhang01,zhang04}, the kinematical evolution of CMEs is divided into a three-phase scenario. The (1) CME initiation starts with a slow rising motion of the duration of some tens of minutes, followed by the (2) impulsive or major acceleration phase, during which the maximum of acceleration and speed is reached. The first two phases clearly happen in the inner corona at distances below 2~Rs \citep{cyr99,vrsnak01CME}, where the maximum acceleration peak is found below $<$0.5~Rs \citep{zhang06,bein11}. As most of CMEs are accompanied by flares \citep{yashiro06}, the energy release of flares is an important parameter for better understanding the driving process of CMEs, e.g.\,by measuring the high energetic hard X-ray (HXR) flux of CME associated flares, it is found that the HXR profile is almost synchronized with the CME acceleration phase indicating a flare-CME feedback relation \citep{temmer08,temmer10}. During the (3) propagation phase into IP space, the CME gets finally adjusted to the speed of the ambient solar medium \citep[e.g.,][]{gopalswamy00}. In this respect the CME mass as well as its cross-section (width) is an important parameter \citep[e.g.,][]{vrsnak10}, and using STEREO data the 3D mass can be calculated \citep[see][]{colaninno09}. From a statistical study, the evolution of CME mass $m$ can be described by Eq.\ref{mass} which shows that the CME mass depends on the observed height $h$ of the CME front, its initially ejected mass $m_0$ (i.e., the proper CME mass ejected from the low corona without occulter effect), and the size of the coronagraph occulter $h_{\rm occ}$ \citep{bein13}. 

\begin{equation}
  \label{mass}
m(h)=m_0\big(1-(h_{\rm occ}/h)^3 \big)+\Delta m(h-h_{\rm occ})
\end{equation}

From this an increase in CME mass is derived with $\Delta$m(h)$\sim$2\%--6\% within a height $h$=10--20~Rs and an initial CME mass of m$_0$ = 10$^{14}$g--10$^{16}$g at a distance $<$ 3~Rs. These results have also important implications for studies on global energetics of flares and CMEs \citep[see e.g.,][]{emslie12}.

\subsection{The early evolution of CMEs}
\begin{figure}
\includegraphics[width=\linewidth]{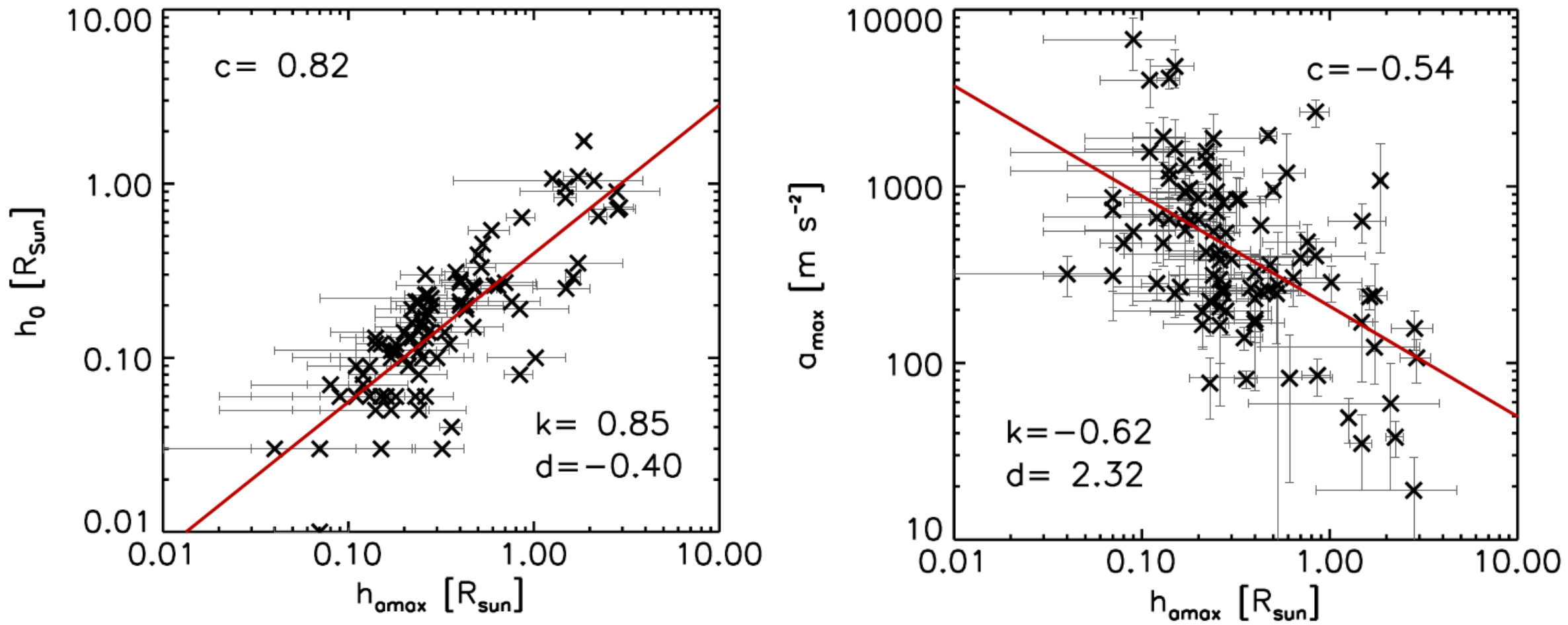}
\caption{Left: Height $h_0$ at which the CME front was first detected against the height at peak acceleration $h_{\rm amax}$. Right: CME peak acceleration $a_{max}$ against $h_{\rm amax}$. Red solid lines are linear regression lines to the data points with parameters $k$, the slope, and $d$, the y-intercept and $c$, the correlation coefficient. Adapted from \cite{bein11}.}
\label{f3}
\end{figure}

Only detailed height-time profiles covering the early evolution of the CME close to the Sun enable us to study the impulsive acceleration phase of a CME. To derive the entire kinematical profile of a CME, usually plane-of-sky measurements are used from coronagraph data allowing views to very low distances from the Sun (e.g., from Mauna Loa observatory), or combined on-disk/off-limb, e.g., in the EUV or SXR wavelength regime, and coronagraph observations \citep[e.g.,][]{vrsnak07,temmer08}. Such studies showed how low in the corona the maximum of the CME acceleration peak is reached \citep[see also][]{gallagher03,zhang06}. A further study using STEREO observations to derive the 3D kinematical profiles of CMEs revealed that the CME reaches its maximum acceleration typically below 2~Rs and that the acceleration profiles of CMEs associated with prominences are different from those without \citep{joshi11}. The statistical study by \cite{bein11} demonstrated that CMEs starting at lower heights also reach their peak acceleration at lower heights and that CMEs that are accelerated at lower heights reach higher peak accelerations (cf.\,Figure~\ref{f3}). A direct consequence of lower CME initiation heights is that the larger magnetic field yields to a stronger Lorentz force and shorter Alfv\'en time scales making the acceleration process more impulsive. Such a scenario is supported from the result that compact CMEs are accelerated more impulsively \citep{vrsnak07-solph}. This clearly shows that characteristic CME properties are set already low in the solar atmosphere.

\subsection{Measuring CMEs in IP space}
For large distances from the Sun ($>$30~Rs) the analysis of CME observational data, e.g., using STEREO heliospheric imagers, becomes more complicated. CMEs are extended and optically thin objects, when observing them at large distances from the Sun, the plane-of-sky assumption is not valid anymore leading to a more complex treatment of geometry and Thomson scattering effects \citep{howard10}. In order to derive the kinematical profiles of CMEs covering the Sun-Earth distance range the measured elongation needs to be converted into radial distance. Several methods have been developed, either using single or multiple spacecraft data, assuming that the CME has a radial outward motion from the source region on the Sun and a certain front geometry (e.g., point-like, circular, elliptic). For CMEs of small width we may use the fixed-$\phi$ approximation \citep[e.g.,][]{sheeley99,rouillard08}. As the method works poorly for wider CMEs \citep{kahler07webb,rouillard09}, the Harmonic-Mean method was proposed by \cite{lugaz09} and \cite{howard09}. Extensions to these methods were developed introducing the fact that CMEs evolve in a self-similar manner \citep[SSE][]{davies12,moestl13}. Other stereoscopic reconstruction methods include Tangent-to-a-sphere \citep[TAS;][]{lugaz10b} or geometric triangulation applied on Heliospheric Imager data \citep[GT;][]{liu10a}, or 3D mask fitting \citep{feng12}. A constraint to the elongation measurements may be given by the additional use of in-situ data, provided the CME is observed remotely and in-situ \citep[see e.g.,][]{rollett12}.

The most important parameter for all of those methods is the propagation direction of a CME. To derive this, a vast amount of techniques was established using different observational data sets from two or more vantage points by different spacecraft. Simple tools, as the tie-point reconstruction or triangulation are applied \citep[e.g.,][]{liu09,maloney09,mierla09,temmer09}. More sophisticated methods cover the forward fitting of a model CME to white light images \citep[see][]{thernisien06,thernisien09,wood09}. By deriving the CME mass from different vantage points, also the 3D propagation direction is obtained \citep{colaninno09,bein11}. Another method applies polarization data and ratio techniques \citep{moran09,deKoning09}. In general, triangulation methods work better for distances close to the Sun, since line-of-sight effects are more effective as the CME expands. A major drawback for most of these methods is the \textit{a priori} assumption of the CME geometry that needs to be applied, hence, errors in kinematics and direction may be large. Even polarization measurements show large scatter when influenced by non-polarized plasma material for e.g.\,the embedded filament \citep[see][]{mierla09}. 

\subsection{Effects of the ambient environment on CME propagation}
The kinematical profiles, as derived with the methods described afore, clearly revealed that the environmental conditions in which a CME is embedded plays a major role in the CME propagation behaviour \citep[e.g.,][]{manchester04}. Close to the Sun, effects of the ambient magnetic field become evident as observational signatures of CME rotation during which most probably an adjustment to the ambient magnetic field structure is taking place \citep[see e.g.,][]{yurchyshyn01,yurchyshyn09,vourlidas11,panasenco13}. Influences of magnetic field pressure gradients are also observed as longitudinal/latitudinal deflection, i.e.\,as non-radial outward motion of the CME when observed close to the Sun \citep[][]{macqueen86,burkepile99,byrne10,foullon11,bosman12,moestl15}. Further out in IP space, the CME propagation is strongly dependent on the interaction with the ambient solar wind \citep[see recent studies by e.g.,][]{savani10,temmer11,rollett14,mays15} and the solar wind drag effect becomes dominant as CMEs which are faster than the solar wind, are slowed down, and those that are slower get accelerated \citep[e.g.,][]{gopalswamy01,vrsnak02}. 

To interpret the kinematical evolution in IP space, the analytical drag based model (DBM) is widely used \citep[][]{vrsnak07zic,vrsnak13,vrsnak14}. The drag acceleration is given by Eq.~\ref{dbm} with $r$ the radial distance of the CME leading edge and $\gamma=c_d A \rho_{\rm SW}/m$. $\gamma$ is the so-called drag parameter depending on $A$ the area of the CME cross-section (as can be derived from 3D reconstructions), $\rho_{\rm SW}$ the ambient density, $m$ the CME mass and c$_d$ the dimensionless drag coefficient \citep[for details see][]{cargill04}. Due to the fact that the background solar wind speed and density may reveal strong gradients such as high speed solar wind streams or other CMEs, the advanced version of the DBM provides a distance dependent solar wind speed as well as density, that can be adjusted to interpret strong changes in the CME kinematics \citep[see][]{zic15}.

\begin{equation}
  \label{dbm}
d^2r/dt^2 = \gamma(r)\big(dr/dt - w(r)\big) \big|dr/dt - w(r)\big|
\end{equation}

The DBM shows that without knowing the spatial distribution of solar wind parameters, we may not be able to fully understand the CME propagation behaviour and consequently to forecast their arrival times and impact speeds at certain distances in IP space. Taking into account the variations in solar activity, the CME occurrence rate ranges from about 0.3 per day (solar minimum activity phase) to 4--5 per day (solar maximum activity phase) \citep[e.g.][]{cyr00}. With transit times of the order of one to five days and propagation speeds of about 500--3000\,km/s, we may expect especially during times of high solar activity, dominating fluctuations due to successive CME eruptions. This preconditioning is a very important aspect and several well observed events could be studied revealing that conditions in IP space are subject to constant change and as such provide different circumstances for every CME. For example, it was found that CMEs ahead may ``clear the way'', making follow-up events super-fast \citep{liu14b,temmer15}. On the other hand, successive CMEs heading towards similar directions, may merge and form complex ejecta of single fronts \citep[e.g.,][]{gopalswamy01,burlaga02,wang02,wu07}. 

The CME-CME merging process is assumed to be revealed in radio observations as strong enhancement in type II bursts \citep[see e.g.,][]{gopalswamy01,gopalswamy02,hillaris11,kahler14}. The effects of such complex CME entities at Earth's magnetosphere cover extended periods of negative $B_z$ \citep[e.g.][]{wang03,farrugia06} making such CMEs more geoeffective and causing more intense geomagnetic storms compared to isolated single events \citep[e.g.][]{burlaga87,farrugia06,xie06,dumbovic15}. Observational evidence for the interaction process itself is hard, if not impossible, to derive e.g.\,the kinematical profiles from remote sensing data point out a strong deceleration starting already hours before the actual CME leading edges are observed to get merged. From this we can infer that a transfer of momentum is taking place \citep[see e.g.,][]{farrugia04,lugaz09,maricic14,mishra15,mishraAsrivas15}. An extremely well documented CME-CME interaction are the events of August 1--2, 2010 which was studied in detail by various authors \citep[][]{harrison12,martinez12,moestl12,liu12,temmer12}. Recent results indicate that the interaction process might be tightly related to the location of the magnetic flux rope, as the slower CME ahead actually presents an MHD obstacle in terms of increased magnetic tension, density, and pressure as well as decreased speed. Since the magnetic flux rope is usually covering a smaller area of the entire CME-shock structure \citep[e.g.,][]{kwon15}, we observe a locally deformed frontal structure for those regions where the flux ropes interact, while other regions, mainly the shock, may freely propagate \citep{temmer14}. In conclusion, the observational data reveal only the consequences of CME-CME interaction, that is not sufficient to fully understand and describe the physics of the merging process.

\begin{figure}
\includegraphics[width=\linewidth]{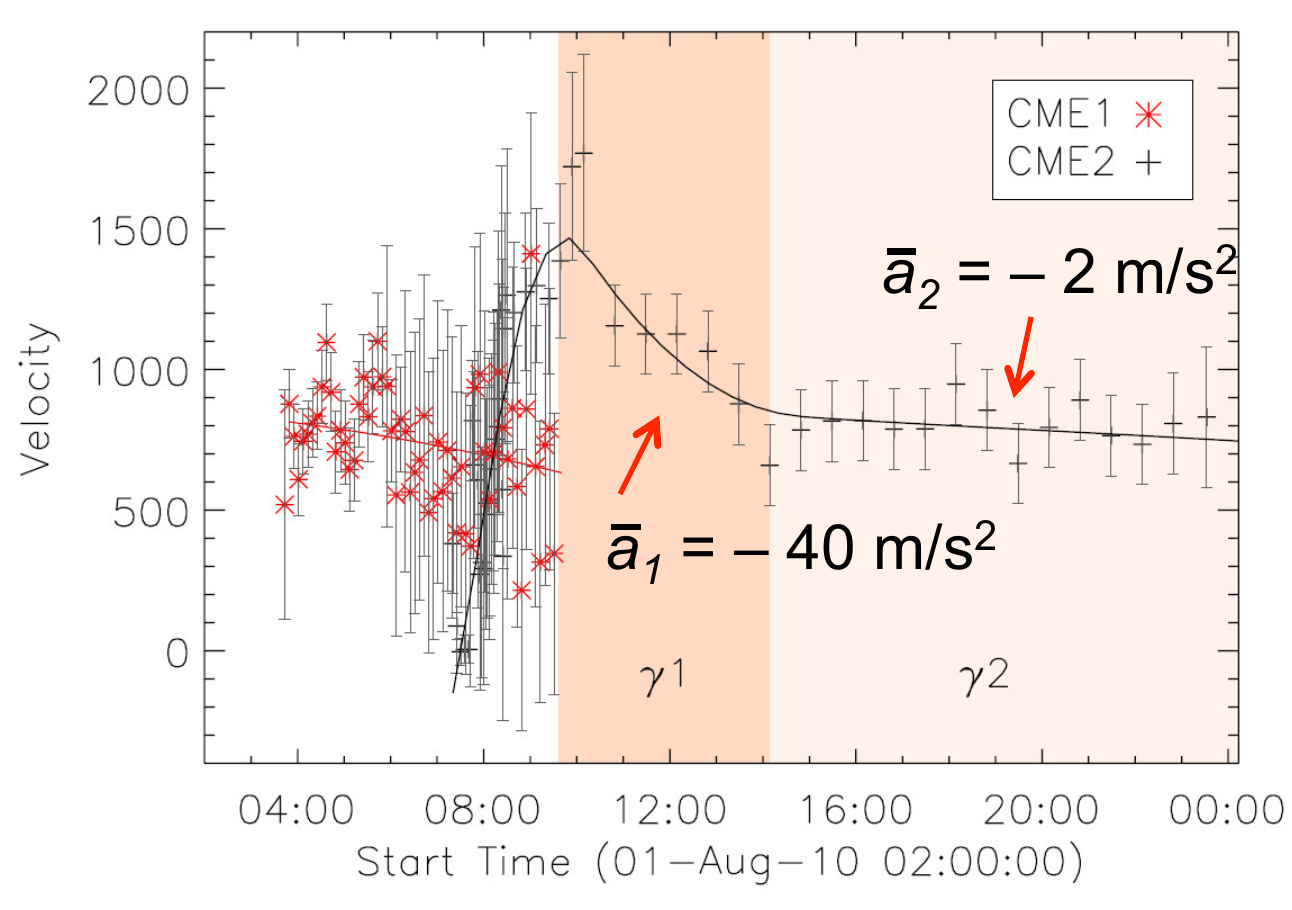}
\caption{Strong changes in the kinematical profiles at far distances from the Sun revealing large deceleration values are derived for interacting CMEs (CME1=CME ahead, CME2=following CME). The CMEs are assumed to propagate further as entity. Adapted from \cite{temmer12}.}
\label{label1}
\end{figure}

\section{Summary and conclusions}
Owing to the wealth of observational data from multiple vantage points starting with the STEREO-era in 2006, CME properties could be studied in more detail than before. From the full CME kinematical profiles, covering in detail their early evolution close to the Sun, it was found that CME properties are set in the low corona. These initial CME properties (speed, acceleration) can be obtained from full-disk observations (e.g., EUV, SXR), and are important parameters to study, as they have an influence on the further propagation behaviour in IP space, and as such, may serve as forecasting proxies \citep[see e.g.,][]{temmer08,bein12,berkebile12}. Especially the ambient magnetic field configuration controls the CME kinematics close to Sun. The CME shape/directivity/speed might be estimated by large-scale magnetic field reconstructions overlying the CME source region \citep[see e.g.,][]{thalmann15}. Therefore, reliably measuring the magnetic field characteristics in the corona from remote sensing data will become the challenge for future missions. 

The methods which were developed to derive the direction of motion of CMEs, deliver crucial input for calculating the radial distance and kinematical evolution of CMEs in IP space from observational data. Using combined observational and model results, the evolution of CMEs can be interpreted in terms of changes of the ambient solar wind parameters. From this it is found that the CME characteristics and the environmental conditions dominate the propagation behaviour and actual transit times of CMEs. For better understanding the physics behind the coupling processes between CMEs and the ambient solar wind, special focus and interest lies in CME-CME interaction events. They are known to form complex magnetic structures producing more strong geomagnetic effects for Earth-directed CMEs than isolated ones, and during times of enhanced solar activity, CME-CME interaction may take place quite often, causing a preconditioning of IP space. Certainly, the increase in magnetic tension forces due to the flux rope interaction plays a key role, and intensifies the drag force of the ambient flow resulting in a strong deceleration of subsequent CMEs. However, it is the lack of plasma and magnetic field parameters at the site where the interaction takes place that hinders us to get a more close insight into the physical processes. The in-situ instruments aboard Solar Orbiter, which will travel at close distances ($<$0.28~AU) to the Sun and Solar Probe Plus, approaching the Sun as close as $\sim$10~Rs (both planned to be launched end 2018) will be of major interest and might give a complementary view on the CME-CME merging processes. More detailed CME-CME interaction studies may greatly improve the modeling of the propagation behaviour of CMEs in IP space.

The constant change and incomplete knowledge of the background conditions to which CMEs are exposed, prevent a major improvement in space weather forecasting. In this respect, models may need a permanent update in almost real-time to simulate the actual solar wind conditions in IP space, finally enabling us to reliably forecast the arrival times and impact speeds of CMEs at Earth.

\acknowledgements
 M.T.\,greatly acknowledges the Fonds zur F\"orderung wissenschaftlicher Forschung (FWF): V195-N16.

%
\bibliographystyle{an}

\end{document}